\documentclass{amsart}
\usepackage[utf8]{inputenc}
\usepackage[foot]{amsaddr}
\usepackage{amsmath}
\usepackage{amsthm}
\usepackage{amsfonts,amssymb}
\usepackage{mathtools}
\usepackage{mathshortcuts}
\usepackage{tikz}
\usepackage{enumitem}
\usepackage[hypertexnames=false]{hyperref}
\usepackage{cleveref}
\usepackage[labelformat=simple,labelfont=normalfont]{subcaption}
\usepackage{autonum}
\usepackage{mleftright}
\usepackage{stmaryrd}

\mleftright

\theoremstyle{plain}
\newtheorem{thm}{Theorem}[section]

\newtheorem{prop}[thm]{Proposition}

\theoremstyle{definition}
\newtheorem{defn}[thm]{Definition}

\crefname{defn}{Definition}{Definitions}
\Crefname{defn}{Definition}{Definitions}
\crefname{thm}{Theorem}{Theorems}
\Crefname{thm}{Theorem}{Theorems}
\crefname{lem}{Lemma}{Lemmas}
\Crefname{lem}{Lemma}{Lemmas}
\crefname{rem}{Remark}{Remarks}
\Crefname{rem}{Remark}{Remarks}
\crefname{prop}{Proposition}{Propositions}
\Crefname{prop}{Proposition}{Propositions}
\crefname{cor}{Corollary}{Corollaries}
\Crefname{cor}{Corollary}{Corollaries}
\crefname{section}{Section}{Sections}
\Crefname{section}{Section}{Sections}
\crefname{figure}{Figure}{Figures}
\Crefname{figure}{Figure}{Figures}
\crefname{equation}{}{}
\Crefname{equation}{}{}
\crefname{enumi}{}{}
\Crefname{enumi}{}{}

\labelcrefformat{subequation}{#2(#1)#3}
\labelcrefrangeformat{subequation}{#3(#1)#4 to #5(#2)#6}
\crefformat{enumi}{#2#1#3}
\crefrangeformat{enumi}{#3#1#4 to #5#2#6}

\makeatletter
\@ifpackageloaded{subcaption}{%
\renewcommand*\subcaption@@label[2]{%
  \@bsphack\begingroup
    \subcaption@ORI@label#1{#2}%
    \let\SK@\@gobbletwo
    \protected@edef\@currentlabel{\csname thesub\@captype\endcsname}%
    \protected@edef\cref@currentlabel{%
      [subs\@captype][\arabic{sub\@captype}][\cref@result]%
      \csname thesub\@captype\endcsname}%
    \subcaption@ORI@label#1{sub@#2}%
  \endgroup\@esphack}%
}
\makeatother

\DeclareMathOperator{\diff}{Diff}
\DeclareMathOperator{\Ceil}{ceil}
\newcommand\sircle{\mathbb{S}}

\title[Approximating Diffeomorphisms by Elements of Thompson's Groups]{Approximating Diffeomorphisms by Elements of Thompson's Groups $F$ and $T$}
\author[Deniz E. Stiegemann]{Deniz E. Stiegemann}
\address{Institut für theoretische Physik, Leibniz Universität Hannover, Appelstraße 2, 30167 Hannover, Germany}
\email{deniz@stiegemann.com}

\begin{document}

\begin{abstract}
We show how to approximate diffeomorphisms of the closed interval and the circle by elements of Thompson's groups $F$ and $T$, respectively. This is relevant in the context of Jones' continuum limit of discrete multipartite systems and its dynamics.
\end{abstract}

\maketitle


\section{Introduction}

Over the past few years, V.\ F.\ R.\ Jones has introduced \emph{discrete} analogues of conformal field theories (CFTs) with the aim of constructing a suitable continuum limit to recover a CFT \cite{Jones17,Jones18,Jones18b}. In the discrete theory, a finitely generated infinite group known as Thompson's group $T$ takes the role of $\diff_+(\sircle^1)$, the group of orientation-preserving diffeomorphisms of the circle. In contrast to diffeomorphisms, the elements of $T$ are piecewise-linear homeomorphisms, which explains the term `discrete'. The idea has already been applied to physics in the context of holography \cite{OsborneStiegemann17}.

The dynamics of the discrete theory is given by (projective) unitary representations of $T$ on an appropriate Hilbert space. While it has been shown that most of these representations are topologically discontinuous and thus unphysical \cite{Jones18,KlieschKoenig18}, interesting exceptions may still exist. The idea -- and challenge -- is to find a procedure that takes a discrete theory as input and then outputs a continuous theory. Such a procedure would certainly include some kind of limit $g_n\to f$, where $g_n\in T$ and $f\in\diff_+(\sircle^1)$.

The purpose of this paper is to clarify how orientation-preserving diffeomorphisms of $\sircle^1$ can be approximated by elements of Thompsons's group $T$. This includes a similar description for orientation-preserving diffeomorphisms of the interval $I=[0, 1]$ and Thompson's group $F$. The corresponding density theorems are certainly known and have been proved for $\operatorname{Homeo}_+(I)$ and $\operatorname{Homeo}_+(\sircle^1)$ in a much more general setting \cite{Zhuang07,BieriStrebel16}. The advantage of our work is a direct proof that is hands-on for the present context and can be directly translated into an algorithm to construct approximations, suitable for the computer.

The reader who is specifically interested in computational applications can find a step-by-step outline of the construction in \cref{sub:outline-of-the-construction}.

\section{Main Facts}
\label{sec:main-facts}

Recall that the dyadic rationals are all numbers of the form $m/2^k$ with $m\in\integers$ and $k\in\naturals=\{0, 1, 2, \dotsc \}$. By a \defemph{breakpoint} of a piecewise linear function we mean the points at which it is not differentiable.

\begin{defn}
  Thompson's group $F$ is the group of piecewise linear homeomorphisms $g$ of the closed unit interval $I=[0, 1]$ such that
  \begin{enumerate}[label=(Th$_\arabic*$)]
    \item\label{item:breakpoints} the breakpoints of $g$ and their images are dyadic rationals;
    \item\label{item:slopes} on intervals of differentiability, the derivatives of $g$ are integer powers of $2$;
  \end{enumerate}
  and Thompson's group $T$ is the group of piecewise linear homeomorphisms $g$ of $\sircle^1$ with these properties.\footnote{These definitions of $F$ and $T$ differ from, but are equivalent to, the standard reference \cite{CannonFloydParry96}. In particular, our definition of $F$ is not minimal since it actually suffices to require that only the breakpoints are dyadic rationals. Their images are then automatically dyadic due to property \cref{item:slopes} and the fact that $0$ is a fixpoint.}
\end{defn}

Let $\diff^1_+(I)$ denote the group of orientation-preserving $C^1$-diffeomorphisms of the interval, and similarly for $\sircle^1$. Our result is stated in terms of the $C^0$-norm
\begin{equation}
  \norm{f}=\sup_x\: \abs{f(x)}.
\end{equation}

\begin{thm}\label{thm:diffapprox}
  For every $f\in\diff^1_{+}(I)$ and $\epsilon>0$, there exists $g\in F$ such that $\norm{f-g}<\epsilon$. Similarly, if $f\in\diff^1_{+}(\sircle^1)$, then there exists $g\in T$ with this property.
\end{thm}

This statement is known and follows from \cite[Thm.~A4.1]{BieriStrebel16} and \cite[Prop.~4.3]{Zhuang07}. It is actually true for all orientation-preserving homeomorphisms. In \cref{sec:approximating-diffeomorphisms} we will give a direct proof of the theorem in the present context.

The next logical question is whether there is an approximation for the first derivatives of diffeomorphisms. While generally elements of both $F$ and $T$ are not everywhere differentiable, we can define a function
\begin{equation}
  d(f, g)=\sup_{x\in \sircle^1\setminus B_g} \abs{f'(x)-g'(x)}.
\end{equation}
that measures the distance between the first derivatives of $f\in\diff^1_+(\sircle^1)$ and $g\in T$ wherever $g'$ is defined. Here $B_g$ denotes the set of breakpoints of $g$. (The definition of $d$ for $\diff^1_+(I)$ and $F$ is analogous.) We can therefore rephrase the question: Given a diffeomorphism $f$ and $\epsilon>0$, is there a function $g$ from the appropriate Thompson group such that $d(f, g)<\epsilon$? The answer is that such an approximation is not possible since the set of all integer powers of $2$ is very sparse in $(0, 1)$. This fact is made precise in the following proposition, which is similar to \cite[Théorème~III.2.3]{GhysSergiescu}.

\begin{prop}\label{prop:discreteness}
  For every $f\in\diff^1_+(\sircle^1)$ which is not a rotation, there exists $\mu>0$ such that $d(f, g)>\mu$ for all $g\in T$. The same holds when $\sircle^1$ is replaced by $I$ and $T$ is replaced by $F$.
\end{prop}

Here the rotations in $\diff^1_+(\sircle^1)$ are all elements $f$ with $f'(x)=1$ for all $x\in \sircle^1$, which includes the identity. In $\diff^1_+(I)$, the identity is the only rotation.

\section{Approximating Diffeomorphisms}
\label{sec:approximating-diffeomorphisms}

In this section, we describe the approximation procedure that represents a proof of \cref{thm:diffapprox}. We begin with a few simplifying observations.

The graph of a piecewise linear function can be described by specifying the (finitely many) breakpoints at which the function is not differentiable, and the images of the breakpoints. For a strictly monotone piecewise linear function $g$, we therefore have a partition of the domain of $g$ by points
\begin{equation}
  x_1 < x_2 < \dotsb < x_n
\end{equation}
and a partition of the codomain of $g$ by the points
\begin{equation}
  g(x_1) < g(x_2) < \dotsb < g(x_n)
\end{equation}
such that $g$ is the function corresponding to the curve of connected line segments through the points
\begin{equation}
  \bigl(x_1, g(x_1)\bigr), \bigl(x_2, g(x_2)\bigr), \dotsc, \bigl(x_n, g(x_n)\bigr).
\end{equation}
In the case of Thompson's groups $F$ and $T$, the breakpoints have to be at dyadic rationals.

\begin{figure}
\begin{subfigure}[b]{0.3\linewidth}
  \begin{tikzpicture}
    \footnotesize

    \def\ri{0.3}
    \def\ro{0.8}

    \draw[thick] (0, 0) circle [radius=\ri];
    \draw[thick] (0, 0) circle [radius=\ro];

    \draw[in=40, in looseness=3, out looseness=0.5] (\ri, 0) to (-\ro, 0) node[left] {$1/2$};
    \draw[bend right=45] (-\ri, 0) to (0, -\ro) node[below] {$3/4$};
    \draw[bend right=45] (0, -\ri) to (\ro, 0) node[right] {$0\cong 1$};
    \node[above] at (0, \ro) {$1/4$};
  \end{tikzpicture}
  \caption{}
  \label{subfig:Texample1}
\end{subfigure}
\begin{subfigure}[b]{0.3\linewidth}
  \begin{tikzpicture}
    \footnotesize

    \def\base{0.7}

    \draw[very thin, color=gray,step=\base] (0, 0) grid (4*\base, 4*\base);

    \draw[->] (0, 0) -- (4.2*\base, 0);
    \draw[->] (0, 0) -- (0, 4.2*\base);

    \node[below left] at (0, 0) {$0$};
    \node[below] at (2*\base, 0) {$1/2$};
    \node[below] at (4*\base, 0) {$1$};

    \node[left] at (0, 2*\base) {$1/2$};
    \node[left] at (0, 4*\base) {$1$};

    \draw (0, 2*\base) -- (2*\base, 3*\base) -- (3*\base, 4*\base);
    \draw (3*\base, 0) -- (4*\base, 2*\base);
  \end{tikzpicture}
  \caption{}
  \label{subfig:Texample2}
\end{subfigure}
\begin{subfigure}[b]{0.3\linewidth}
  \begin{tikzpicture}
    \footnotesize

    \def\base{0.7}

    \draw[very thin, color=gray,step=\base] (0, 0) grid (4*\base, 6*\base);

    \draw[->] (0, 0) -- (4.2*\base, 0);
    \draw[->] (0, 0) -- (0, 6.2*\base);

    \node[below left] at (0, 0) {$0$};
    \node[below] at (2*\base, 0) {$1/2$};
    \node[below] at (4*\base, 0) {$1$};

    \node[left] at (0, 2*\base) {$1/2$};
    \node[left] at (0, 4*\base) {$1$};
    \node[left] at (0, 6*\base) {$3/2$};

    \draw (0, 2*\base) -- (2*\base, 3*\base) -- (3*\base, 4*\base) -- (4*\base, 6*\base);
  \end{tikzpicture}
  \caption{}
  \label{subfig:Texample3}
\end{subfigure}
\caption{Three representations of the same element of Thompson's group $T$: \labelcref{sub@subfig:Texample1} as a map $\sircle^1\to \sircle^1$, here drawn by indicating how breakpoints (on the inner circle) are mapped to their images (on the outer circle); \labelcref{sub@subfig:Texample2} the usual representation as a function $[0, 1]\to [0, 1]$; \labelcref{sub@subfig:Texample3} the representation as a function $[0, 1]\to\reals$, which we will use -- note that it is a homeomorphism onto its image $[1/2, 3/2]$.}
\label{fig:Texample}
\end{figure}

Given any homeomorphism $f\colon \sircle^1\to \sircle^1$, we can identify it with a homeomorphism $\tilde f\colon\reals\to\reals$ that satisfies
\begin{equation}
  \tilde f(x+1)=\tilde f(x)+1.
\end{equation}
In particular, $\tilde f|_{[0, 1]}$ is continuous, which will be needed later. An example is shown in \cref{fig:Texample}.


\subsection{Outline of the Construction}
\label{sub:outline-of-the-construction}


Before we come to technical details, we present a rough outline of the proof for the case of $\diff^1_+(I)$ and Thompson's group $F$. Let $f\in\diff^1_+(I)$ be given.
\begin{enumerate}
  \item Divide the domain of $f$ into $n$ small intervals of equal length, where $n$ is a power of $2$. Therefore the breakpoints $\xi_i$ of the partition are dyadic rationals.
  \item For each breakpoint $\xi_i$ choose a dyadic rational $\eta_i$ close to the image $f(\xi_i)$.
  \item Find a piecewise linear homeomorphism $\gamma_i\colon[\xi_i, \xi_{i+1}]\to [\eta_i, \eta_{i+1}]$ for each $i = 0, \dotsc, n-1$ that serves as a \defemph{dyadic interpolation} from the point $(\xi_i, \eta_i)$ to the point $(\xi_{i+1}, \eta_{i+1})$, which means that $\gamma_i$ has breakpoints at dyadic rationals and its slopes are powers of $2$ (\cref{sub:dyadic-interpolation}).
\end{enumerate}
By defining a function $g\colon[0, 1]\to [0, 1]$ whose values on the interval $[\xi_i, \xi_{i+1}]$ are determined by $\gamma_i$, we obtain a homeomorphism $g\in F$ close to $f$.

\subsection{Dyadic Interpolation}
\label{sub:dyadic-interpolation}

\begin{figure}
  \begin{tikzpicture}
    \def\base{0.4}
    \draw (0, 0) rectangle (2*8*\base, 11*\base);
    \foreach \y in {1, ..., 10}
      \draw[very thin, color=gray, yshift=\y*\base cm] (0, 0) -- (2*8*\base, 0);
    \draw (0, 0) node[below left] {$0$};
    \draw (0, 1*\base) node[left] {$\frac{1}{2^6}$};
    \draw (0, 2*\base) node[left=6pt] {$\frac{2}{2^6}$};
    \draw (0, 3.8*\base) node[left=5pt] {$\vdots$};
    \draw (0, 11*\base) node[left] {$\frac{11}{2^6}$};

    \draw (1*8*\base, 0) node[below] {$\frac{1}{2^3}$} -- ++(0, 11*\base);
    \draw (2*8*\base, 0) node[below] {$\frac{2}{2^3}$};

    \fill[lightgray!50] (0, -0.65) rectangle ++(21.6*\base, -0.6);
    \fill[lightgray!10] (0, -1.25) rectangle ++(21.6*\base, -0.6);
    \fill[lightgray!50] (0, -1.85) rectangle ++(21.6*\base, -0.6);
    \foreach \x in {1, 3}
      \draw[xshift=\x*4*\base cm, thin] (0, 11*\base) -- (0, -0.6) node[below] {$\frac{\x}{2^4}$};
    \foreach \x in {1, 3, 5, 7}
      \draw[xshift=\x*2*\base cm, thin] (0, 11*\base) -- (0, -1.2) node[below] {$\frac{\x}{2^5}$};
    \foreach \x in {1, 3, 5}
      \draw[xshift=\x*1*\base cm, thin] (0, 11*\base) -- (0, -1.8) node[below] {$\frac{\x}{2^6}$};

    \draw (2*8*\base+0.5, -0.95) node[right] {$n=1$};
    \draw (2*8*\base+0.5, -1.55) node[right] {$n=2$};
    \draw (2*8*\base+0.5, -2.15) node[right] {$n=l=3$};

    \draw[thick, line cap=round] (0, 0) -- (3*2*\base, 3*2*\base) -- (2*8*\base, 11*\base);
  \end{tikzpicture}
  %
  %
  %
  %
  \caption{Illustration of how to cut the sides of a dyadic rectangle such that all sides are divided into dyadic partitions with equally many subintervals. In this example, $m_1/2^{k_1}=11/2^6$ and $m_2/2^{k_2}=2/2^3$. Since $11>2$, we divide the left side of the rectangle into $11$ intervals, each of length $1/2^6$. The bottom side is first divided into $2$ intervals, each of length $1/2^3$. Then we successively cut all its intervals in half, repeatedly going from left to right, until the bottom side is also divided into $11$ intervals. The thick line shows the graph of the piecewise linear function arising from these partitions.}
  \label{fig:partitions}
\end{figure}

%
Let two distinct points $p=(p_1, p_2)$ and $q=(q_1, q_2)$ in $\reals^2$ be given, with $p_1<q_1$ and $p_2<q_2$ and such that all coordinates $p_i$, $q_i$ are dyadic rational numbers. Then $r=q-p$ also has dyadic rational coordinates $r_1$ and $r_2$ which can be written as
\begin{equation}
  r_1=\frac{m_1}{2^{k_1}}, \quad r_2=\frac{m_2}{2^{k_2}}
\end{equation}
with $m_1, m_2, k_1, k_2\in\naturals$ and $m_1, m_2>0$. We proceed as illustrated in \cref{fig:partitions}. Let $(a, b)=(1, 2)$ if $m_1\le m_2$ and $(a, b)=(2, 1)$ if $m_1>m_2$, so that $m_b=\max\{m_1, m_2\}$ and $m_a=\min\{m_1, m_2\}$. Set $d=m_b-m_a$. For the moment, assume $d>0$. Consider the sequence $(c_n)$ defined by $c_0=0$ and
\begin{equation}
  c_n=m_a\sum_{i=0}^{n-1} 2^i=m_a(2^n-1)
\end{equation}
for $n\ge 1$. Let $l\ge 1$ be the smallest integer with $c_l\ge d$. Define a sequence of dyadic numbers $\xi_1, \dotsc, \xi_d$ by setting
\begin{equation}
  \xi_{i+c_n}=\frac{2i-1}{2^{k_a+n}}
\end{equation}
for all $i, n$ with either $1\le i\le 2^nm_a$ and $0\le n\le l-1$, or $1\le i\le d-c_{l-1}$ when $n=l$. Set
\begin{equation}
  X=\Setcond{\frac{m}{2^{k_a}}}{0\le m\le m_a} \cup \{\xi_1, \dotsc, \xi_d\}
\end{equation}
for $d>0$ and
\begin{equation}
  X=\Setcond{\frac{m}{2^{k_a}}}{0\le m\le m_a}
\end{equation}
for $d=0$. We arrange the $m_a+d=m_b$ elements of $X$ in increasing order and denote them $x^a_1\le\dotsb\le x^a_{m_b}$. They are the breakpoints of a standard dyadic partition of $[0, m_a/2^{k_a}]$ into $m_b$ intervals. Furthermore, set $x^b_m=m/2^{k_b}$ for $0\le m\le m_b$. The points
\begin{equation}
  p_1 + x^1_1,\ p_1+ x^1_2,\ \dotsc,\ p_1 + x^1_n
\end{equation}
and
\begin{equation}
  p_2 + x^2_1,\ p_2+ x^2_2,\ \dotsc,\ p_2 + x^2_n
\end{equation}
form standard dyadic partitions dividing the intervals $[p_1, p_2]$ and $[q_1, q_2]$, respectively, into equally many subintervals. To these partitions corresponds a piecewise linear function. By construction, it is bijective, has breakpoints only at dyadic rationals, and only slopes wich are powers of $2$.

\subsection{Finding Dyadic Rationals}

Let $0<p<q$ be given. Since the dyadic rationals are dense in $\reals$, one can always find a dyadic number in the open interval $(p, q)$. For an example, let
\begin{equation}
  \overline\Ceil(x) = \min\setcond{n\in\integers}{n>x}=\begin{cases}
    x+1& \text{if $x\in\integers$},\\
    \ceil{x}& \text{otherwise}.
  \end{cases}
\end{equation}
Set
\begin{gather}
  k = \max\left\{0, \overline\Ceil(-\log_2 (q-p))\right\},\\
  m= \overline\Ceil (2^k p).
\end{gather}
Then $m, k\in\naturals$, and $m/2^k\in (p, q)$ is a dyadic rational.

\subsection{The Construction}

We proceed with the construction of approximations, which then proves \cref{thm:diffapprox}. Let $f\in\diff^1_+(I)$ and $\epsilon>0$ be given, and assume $\epsilon<1$ without loss of generality. Set $S=\max_{x\in I} f'(x)$ and note that $S\ge 1$. Let $\Delta=\lceil-\log_2\frac{\epsilon}{3S}\rceil\in\naturals$ and $n=2^\Delta$, and note that $\Delta\ge 1$. Set
\begin{equation}
  \xi_i=i/n, \quad i=0, \dotsc, n.
\end{equation}
(This implies that $\xi_0=f(\xi_0)=0$ and $\xi_n=f(\xi_n)=1$.)
Moreover, set
\begin{equation}
  \delta=\min\{\epsilon/2, (f(\xi_n)-f(\xi_{n-1})/2)\}
\end{equation}
and note that the interval
\begin{equation}
  I_i=(\max\{f(\xi_{i-1})+\delta, f(\xi_i)\},f(\xi_i)+\delta)
\end{equation}
is non-empty and a subset of $(0, 1)$ for $i=1, \dotsc, n-1$. We pick a dyadic rational $\eta_i\in I_i$ for each $i=1, \dotsc, n-1$. Let $\eta_0=0$ and $\eta_n=1$, and define the function $g\colon [0, 1]\to [0, 1]$ by setting
\begin{equation}\label{eq:gdefinition}
  g(x)=\gamma_i(x)
\end{equation}
for $x\in [\xi_i, \xi_{i+1}]$ and $i=0, \dotsc, n-1$, where $\gamma_i$ is a dyadic interpolation from the point $(\xi_i, \eta_i)$ to the point $(\xi_{i+1}, \eta_{i+1})$. From the definitions of $\gamma$, $\{\xi_i\}$ and $\{\eta_i\}$ it is clear that $g\in F$. Furthermore, for all $i=0, \dotsc, n-1$ and $x\in [\xi_i, \xi_{i+1}]$, consider the sequence of statements
\begin{align}
  \abs{g(x)-f(x)}&\le g(\xi_{i+1})-f(\xi_i) \label{eq:epsilon1}\\
  &< f(\xi_{i+1})-f(\xi_i)+\epsilon/2 \label{eq:epsilon2}\\
  &= \frac{f(\xi_{i+1})-f(\xi_i)}{\xi_{i+1}-\xi_i} (\xi_{i+1}-\xi_i)+\epsilon/2 \label{eq:epsilon3}\\
  &< S2^{-\Delta}+\epsilon/2 \label{eq:epsilon4}\\
  &< \epsilon/3+\epsilon/2 < \epsilon \label{eq:epsilon5}.
\end{align}
\cref{eq:epsilon1} holds since $f$ and $g$ are strictly increasing and $g(\xi_i)>f(\xi_i)$. For $\cref{eq:epsilon2}$, recall that $g(\xi_{i+1})=\eta_{i+1}<f(\xi_{i+1})+\delta$. \cref{eq:epsilon3,eq:epsilon4,eq:epsilon5} are obvious. We have thus found $g\in F$ with $\max_{x\in [0, 1]}\abs{f(x)-g(x)}<\epsilon$.

If instead $f\in\diff^1_{+}(\sircle^1)$, $f$ corresponds to a function $\tilde f\colon\reals\to\reals$ with $\im(\tilde f)=[u, u+1]$ for some $u\in\reals$ and such that $\tilde f\colon[0, 1]\to [u, u+1]$ is a diffeomorphism (as explained above). Define $S$, $\Delta$, $n$, $\xi_i$ and $I_i$ as above, but with $\delta=\min\{\epsilon/2, (\tilde f(\xi_1)-\tilde f(\xi_0))/2\}$. Choose $\eta_i\in I_i$ for $i=1, \dotsc, n-1$ as before. Let $\eta_0$ be a dyadic rational in the interval $(\tilde f(\xi_0)+\delta, \tilde f(\xi_1))$ and set $\eta_n=\eta_0+1$. This ensures that
\begin{equation}
  \max\{\tilde f(\xi_{n-1})+\delta, \tilde f(\xi_n) \}<\eta_n.
\end{equation}
Now we can define a function $\tilde g\colon [0, 1]\to\reals$ as in \cref{eq:gdefinition}. It follows that \cref{eq:epsilon1,eq:epsilon2,eq:epsilon3,eq:epsilon4,eq:epsilon5} hold, and that $g\in T$ upon taking the quotient $\sircle^1=\reals/\integers$. This concludes the proof of \cref{thm:diffapprox}. \qed

\section{$C^1$-Discreteness}
\label{sec:c1-discreteness}

Finally, we show that it is not possible to go beyond $C^0$-approximation.
Note that the proof is also valid in the more general case when $\diff^1_+(I)$ and $\diff^1_+(\sircle^1)$ are replaced by the sets of all differentiable bijections of $I$ or $\sircle^1$, respectively, whose inverses are also differentiable.

\begin{proof}[Proof of \cref{prop:discreteness}]
  Let $g\in T$ and $f\in\diff^1_+(\sircle^1)$. We will identify $f$ and $g$ with functions on the interval $[0, 1]$ as before. Let $x_0\in [0, 1]\setminus B_{g}$. The two powers of $2$ closest to $f'(x_0)$ are given by
  \begin{equation}
    2^{\floor{\log_2 f'(x_0)}} \le f'(x_0) \le 2^{\ceil{\log_2 f'(x_0)}}.
  \end{equation}
  If $f'(x_0)$ is not a power of $2$, the inequalities are strict and therefore
  \begin{equation}
    d(f, g) \ge \min\Bigl\{ \bigl\lvert f'(x_0)-2^{\floor{\log_2 f'(x_0)}}\bigr\rvert, \bigl\lvert f'(x_0)-2^{\ceil{\log_2 f'(x_0)}}\bigr\rvert \Bigr\}>0.
  \end{equation}
  The case that $f'(x_0)$ is not a power of $2$ for some $x_0\in [0, 1]\setminus B_g$ occurs for all differentiable $f\in\diff^1_+(\sircle^1)$ except for rotations. For if $f$ is not a rotation, there exists $x_1\in [0, 1]$ with $f'(x_1)=c\neq 1$. By the mean value theorem, there also exists $x_2\in [0, 1]$ with $f'(x_2)=1$. Without loss of generality, assume $c<1$ and $x_1<x_2$. Then by Darboux's theorem, $[c, 1]\subset f'([x_1, x_2])$. Since $B_g$ is finite, $[c, 1]\setminus f'(B_g)\subset\im(f')$ surely contains points which are not powers of $2$.

  It is clear that $\diff^1_+(I)$ and $F$ are a special case of this argument, which concludes the proof.
\end{proof}

\section*{Acknowledgements}

I would like to thank Tobias Osborne for introducing me to the problem and many helpful discussions. I am also grateful to Terry Farrelly and Ramona Wolf for numerous comments and a careful reading of the manuscript.

This work was supported by the DFG through SFB 1227 (DQ-mat) and the RTG 1991, the ERC grants QFTCMPS and SIQS, and the cluster of excellence EXC201 Quantum Engineering and Space-Time Research.


\bibliographystyle{halpha}
\bibliography{lit}

\begin{thebibliography}{Jon18b}

\bibitem[BS16]{BieriStrebel16}
Robert Bieri and Ralph Strebel.
\newblock On {{Groups}} of {{PL}}-homeomorphisms of the {{Real Line}}.
\newblock volume 215 of {\em Math. Surveys Monogr.} Amer. Math. Soc., 2016.

\bibitem[CFP96]{CannonFloydParry96}
J.~W. Cannon, W.~J. Floyd, and W.~R. Parry.
\newblock Introductory notes on {{Richard Thompson}}'s groups.
\newblock {\em Enseign. Math.}, 42:215--256, 1996.

\bibitem[GS87]{GhysSergiescu}
E.~Ghys and V.~Sergiescu.
\newblock Sur un groupe remarquable de difféomorphismes du cercle.
\newblock {\em Comment. Math. Helv.}, 62(1):185--239, 1987.

\bibitem[Jon17]{Jones17}
V.~F.~R. Jones.
\newblock Some unitary representations of {{Thompson}}’s groups {{$F$}} and
  {{$T$}}.
\newblock {\em J. Comb. Algebra}, 1(1):1--44, 2017.

\bibitem[Jon18a]{Jones18}
V.~F.~R. Jones.
\newblock A {{No}}-{{Go}} {{Theorem}} for the {{Continuum}} {{Limit}} of a
  {{Periodic}} {{Quantum}} {{Spin}} {{Chain}}.
\newblock {\em Commun. Math. Phys.}, 357(1):295--317, 2018.

\bibitem[Jon18b]{Jones18b}
V.~F.~R. Jones.
\newblock Scale invariant transfer matrices and {{Hamiltionians}}.
\newblock {\em J. Phys. A: Math. Theor.}, 51(10):104001, 2018.

\bibitem[KK]{KlieschKoenig18}
A.~Kliesch and R.~König.
\newblock Continuum limits of homogeneous binary trees and the {{Thompson}}
  group.
\newblock arXiv:1805.04839.

\bibitem[OS]{OsborneStiegemann17}
T.~J. Osborne and D.~E. Stiegemann.
\newblock Dynamics for holographic codes, arXiv:1706.08823.

\bibitem[Zhu08]{Zhuang07}
D.~Zhuang.
\newblock Irrational stable commutator length in finitely presented groups.
\newblock {\em J. Mod. Dyn.}, 2(3):499--507, 2008.

\end{thebibliography}

\end{document}